\documentclass{article}
\usepackage[utf8]{inputenc}
\usepackage[T2A]{fontenc}
\usepackage{authblk}

\usepackage{slashed}

\usepackage{amsmath}
\usepackage{mathtools}

\usepackage{graphicx}
\usepackage[left=2cm,right=2cm, top=2cm,bottom=2cm,bindingoffset=0cm]{geometry}
\usepackage[usenames]{color}
\usepackage{colortbl}
\usepackage{indentfirst}
\usepackage{verbatim}
\usepackage{subcaption}
\usepackage[colorlinks=red, urlcolor=blue]{hyperref}
\usepackage{comment}
\usepackage{physics}
\usepackage[normalem]{ulem}
\usepackage{natbib}

\usepackage{mathtools}

\bibpunct{[}{]}{,}{n}{,}{,}

\usepackage{amssymb}
\usepackage{bm}
\usepackage{xcolor}
\usepackage{hyperref}

\everymath{\displaystyle} 

\title{Absorption of a twisted photon by an electron in strong magnetic field}
\author{A.A. Shchepkin\thanks{a.shchepkin@metalab.ifmo.ru}}
\author{D.V. Grosman\thanks{dmitriy.grosman@metalab.ifmo.ru}}
\author{I.I. Shkarupa} 
\author{D.V. Karlovets\thanks{dmitry.karlovets@metalab.ifmo.ru}}
\affil{{\small School of Physics and Engineering, ITMO University, 197101, St. Petersburg, Russia}}
\begin{document}

\maketitle

\begin{abstract}
   The work investigates absorption of a twisted photon, which possesses quantized total angular momentum (TAM), by a relativistic electron with the Lorentz factor $\gamma \sim 1-10$ in a strong magnetic field up to the Schwinger limit, $H_c = 4.4\cdot 10^{13}$ G. We examine the absorption cross sections and their dependence on the parameters of the incident photon and the initial Landau electron. It is found that total absorption cross sections decrease as angular momentum of the incident photon increases and increase as angular momentum of the initial electron grows. The process is also compared across different magnetic field strengths, and the contribution of various electron spin transitions to total absorption cross section is analyzed. We also find that the processes without an electron spin flip dominate and, on top of that, an asymmetry in the ``spin-down'' $\rightarrow$ ``spin-up'' and the ``spin-up'' $\rightarrow$ ``spin-down'' transitions is observed. Specifically, the cross sections for the ``spin-down'' $\rightarrow$ ``spin-up'' transition are larger, which can be interpreted as an analogy of the Sokolov-Ternov effect present for photon emission. Finally, the cross sections are found to be almost constant as the transverse momentum of the photon varies from $0.1$ eV to $100$ keV. Our findings can help to improve the understanding of the QED processes in critical fields, typical for astrophysical environments, e.g. magnetospheres of neutron stars. 
\end{abstract}


\section{Introduction}
It has recently been argued that an electron can emit an electromagnetic field carrying orbital angular momentum (OAM), when moving in a helical path, for example in an external magnetic field \cite{Epp2023, Bordovitsyn2012Jun}, a helical undulator \cite{katoh2017helical, Matsuba2018Jul} or a circularly-polarized laser wave \cite{Hemsing2011Apr, Zhang2021Jul}. The emitted field consists of the so-called twisted (or vortex) photons \cite{allen1992orbital, andrews2012angular, torres2011twisted, SerboUFN}. The quantum description of the photon emission is an intricate problem due to entanglement of the final particles and depending on the post-selection protocol the emitted photons may or may not posses a definite OAM \cite{EPJC, short_EPJC}. Moreover, the traditional description of the radiation processes implies that both the final electron and the photon are detected. However, recently there have been several works, where the quantum state of the photon emitted by an electron in a magnetic field was derived regardless of its detection scheme \cite{Karlovets2023Sep, Pavlov2024}. It was established that an electron transition between different Landau states (in a magnetic field) results in the emission of vortex photons. 

The Landau states of charged particles in an external magnetic field are the basis of the quantum theory of syncrotron radiation \cite{sokolov1986radiation}, which plays a key role in the description of astrophysical objects \cite{bordovitsyn1999synchrotron}, the operation of modern storage rings, and the production of spin-polarized electron beams. These states also find numerous applications in other areas including electron microscopy \cite{schattschneider2014imaging}, physics of plasma \cite{Shah} and solid state physics \cite{Sadovsky, Li, Yang, LLsplitting} and are of the crucial importance for the quantum dynamics of charged particles in Penning traps \cite{Larson_1988, Kluge_2003}. 

A realistic experiment or a physical process where radiation occurs can hardly ever be limited to a single emission process. In perturbative QED the emission of a higher number of photons is suppressed by the fine structure constant $\alpha \approx 1/137$. However, provided a large density of charged particles in an external field absorption an re-emission processes come into play. This is the case, for example, for magnetospheres of neutron stars where the density of the charge carriers can vary within the range $10^{12} - 10^{24}\, \text{cm}^{-3}$ \cite{Goldreich1969Aug, Mushtukov2022May}. Inspired by the investigations of the photon emission by an electron in a magnetic field (see e.g. \cite{Karlovets2023Sep}, \cite{Pavlov2024}), we aim to consider the reverse scenario. In this paper we study the absorption process in a strong magnetic field up to the critical value $H_{\rm{c}} = m_{\rm{e}}^2c^3/|e|\hbar \approx 4.412\times 10^{13}$ G which can be found in an astrophysical environment (e.g., near the neutron star poles \cite{Shapiro1983Jul, Kaspi2017Aug}).

Photon absorption by electrons in the strong magnetic field of an astrophysical environment has been considered before \cite{Ginzburg1969Sep, Daugherty1978Aug, Harding1991Jun}. However, the theoretical description of the process was performed for photons with a definite momentum (plane waves). Since it has been shown that radiation generated by electrons in a magnetic field is generally twisted \cite{Karlovets2023Sep, Pavlov2024}, it is of interest to generalize the theoretical description of absorption processes to account for photon angular momentum.

The paper is organized as follows. In Sec. \ref{sec:wfs} we discuss the relativistic Landau states of an electron in a constant and uniform magnetic field and in Sec. \ref{sec:ph} is introduced the vector potential of the incident vortex photon. In Sec. \ref{sec:mat_elem} we derive the first-order S-matrix element for photon absorption. In Sec. \ref{sec:kinem} the kinematics of the process is discussed. In Sec. \ref{sec:results} the obtained results are analyzed and Sec. \ref{sec:disc_n_conc} is dedicated to discussion and conclusion. Throughout the paper we use the natural system of units with $\hbar=c=1$ and the electron mass and charge are denoted as $m_e$ and $e$ ($e < 0$), respectively. The Coulomb gauge is used for the potential of the incident photon.

\section{Relativistic Landau states}
\label{sec:wfs}
In a constant and uniform magnetic field $\bm{H} = \{0,0,H\}$ a charged particle (an electron, from hereon) is described by the Dirac equation
\begin{equation}
\label{Land}
    \left(\gamma^{\mu}(i\partial_{\mu} - e\mathcal{A}_{
    \mu})
     -m_e\right)\Psi(x) = 0.
\end{equation}
In the symmetric gauge, corresponding to the following vector potential of the magnetic field
\begin{equation}
    \mathcal{A}^{\mu} = \frac{H}{2}\{0,-y,x,0\},
\end{equation}
the solutions to \eqref{Land} can be chosen in the from of the well known Landau states with a definite value of total angular momentum (TAM), comprising of spin and orbital contributions, which are described by Laguerre-Gaussian functions \cite{ciftja2020detailed}. Alternatively, one can choose a different gauge for the vector potential $\mathcal{A}^{\mu} = H\{0,0,y,0\}$ and obtain a basis of solutions in the form of Hermite-Gaussian functions \cite{landau1989diamagnetismus}, which are also commonly referred to as Landau states. In this work, we will use the Laguerre-Gaussian packets as we consider the initial electron to be possessing the TAM. Moreover, these states are normalized in both transverse directions (relative to the particle propagation axis) and allow for a simpler and clearer analysis of the TAM transfer. They can be constructed from the scalar wave functions that satisfy the following equation
\begin{equation}
\label{KG}
    \left[(i\partial_{\mu} - e\mathcal{A}_{\mu})^2 - m_e^2 + eH \Sigma_z\right]\Phi(x) = 0,
\end{equation}
where
\begin{equation}
    \Sigma_z = \begin{pmatrix}\sigma_z & 0\\0&\sigma_z\end{pmatrix} = \text{diag}\{1,-1,1,-1\}.
\end{equation}
One can see \cite{Landau_4} for the detailed derivation of \eqref{KG}. The solution to \eqref{KG} can be presented as the following:
\begin{equation}
\label{sol}
    \Phi(x) = \begin{pmatrix}
        c_1 \Phi_{+}(x)\\
        c_2 \Phi_{-}(x)\\
        c_3 \Phi_{+}(x)\\
        c_4 \Phi_{-}(x)
    \end{pmatrix}, \;\; \Phi_{\pm}(x) = N_{s,j\mp 1/2} \psi_{s,j \mp 1/2}(\tilde\rho)e^{-it\varepsilon_{s,j} + i(j \mp 1/2)\varphi + ip_z z}, \;\; \psi_{s,l}(\tilde\rho) = \tilde\rho^{|l|}L_s^{|l|}(\tilde\rho^2)e^{-\tilde\rho^2/2}.
\end{equation}
Here $s = 0, 1, 2, ..., j = \pm 1/2, \pm 3/2, ...$ are the radial quantum number of the electron and its TAM projection to the magnetic field respectively, $\tilde\rho = \rho/\rho_H$, where 
\begin{equation}
    \rho_H  =  \sqrt{2/|e|H}
\end{equation}
is the coherence length of the ground Landau state. The energy of the electron in a Landau state is related to the quantum numbers via
\begin{equation}
    \varepsilon_{s,j}^2 = m_e^2 + p_z^2 + 4/\rho_H^2(s + j + 1/2),
\end{equation} 
where $p_z$ is the longitudinal momentum of the electron. Now, to obtain the solution to the Dirac equation \eqref{Land} we have to act on \eqref{sol} with the projection operator
\begin{equation}
\label{gen}
    \Psi(x) = \left[\gamma^{\mu}(i\partial_{\mu} - e\mathcal{A}_{\mu}) + m_e\right]\Phi(x).
\end{equation}
The Dirac equation and the corresponding solutions account for the electron spinor degrees of freedom. However, spin and angular momentum separately are not proper quantum numbers in a relativistic theory as it is only the TAM that is a conserved quantity. Thus, there is an ambiguity as to how TAM is separated into spin and OAM constituents,
\begin{equation}
    \hat{j}_z = \hat{l}_z + \frac{1}{2}\hat{\Sigma}_z.
\end{equation}
The general solution \eqref{gen} is an eigenfunction of the TAM operator, but depending on the coefficients $c_i$ its spinor structure is different. Since the
discussion of spin-related phenomena is not the aim of this work (see for details \cite{ST1974, Aleksandrov2020Aug, Bauke2014May}), let us choose two orthogonal spin states obtained by either setting $c_{i = 1} = 1, c_{i \ne 1} = 0$ or $c_{i = 2} = 1, c_{i \ne 2} = 0$ in \eqref{sol}. Such a choice is in accordance with the previous work \cite{vanKruining2017Jul} and also with \cite{Pavlov2024}, where different solutions of the Dirac equation in a magnetic field were examined. The explicit expressions for the electron bispinors are
\begin{equation} \label{eq:bisp_spin_up} 
    \Psi_\uparrow = N_{s,j-1/2}\begin{pmatrix}
        (\varepsilon + m_e)\psi_{s,j-1/2}(\tilde\rho)e^{-i\varphi/2} \\
        0 \\
        p_z \psi_{s,j-1/2}(\tilde\rho)e^{-i\varphi/2} \\
        2\rho_H^{-1}i \psi_{s,j+1/2}(\tilde\rho)e^{i\varphi/2}
    \end{pmatrix} e^{i(p_z z + j\varphi - \varepsilon t)},
\end{equation}
\begin{equation} \label{eq:bisp_spin_down}
    \Psi_\downarrow = N_{s,j+1/2}\begin{pmatrix}
        0 \\
        (\varepsilon + m_e)\psi_{s,j+1/2}(\tilde\rho)e^{i\varphi/2} \\
        -2\rho_H^{-1}i\left(s + j + \frac{1}{2}\right)\psi_{s,j-1/2}(\tilde\rho)e^{-i\varphi/2} \\
        -p_z \psi_{s,j+1/2}(\tilde\rho)e^{i\varphi/2}
    \end{pmatrix} e^{i(p_z z + j\varphi - \varepsilon t)}.
\end{equation}
The indices $\uparrow$ and $\downarrow$ stand for the states where the electron spin $z$-projection, $\sigma_z$, equals $1/2$ and $-1/2$ respectively. The normalization constants are determined from $\int \Psi^{\dagger}\Psi dV = 1$ and are obtained to be
\begin{equation} \label{eq:e_norm_const}
    N_{s,j\mp 1/2} = \sqrt{\frac{s!}{2\pi L(s + |j \mp 1/2|)!\varepsilon(\varepsilon + m_e) \rho_H^2}},
\end{equation}
with $L$ being the normalization length. As mentioned before, the states \eqref{eq:bisp_spin_up}, \eqref{eq:bisp_spin_down} posses a definite value of TAM, and in the limit of a weak magnetic field $H \ll H_c$ they become the eigenstates of the Pauli spin operator 
\begin{equation}
    \hat{\bm{s}} = \frac{1}{2}\bm{\sigma}, \;\; \bm{\sigma} = (\sigma_x,\sigma_y,\sigma_z),
\end{equation}
where $\sigma_i$ are the Pauli matrices, with the eigenvalues $\pm 1/2$. For this reason, we will refer to them as ``spin-up'' and ``spin-down'' states, respectively.

\section{Vector potential of the vortex photon}
\label{sec:ph}
Recently, it has been shown that an electron transition between different Landau states in a magnetic field results in the emission of a vortex photon having the form of a Bessel beam, which is described in detail in \cite{SerboUFN}. Such a photon state is characterized by a definite energy, $\omega$, longitudinal momentum, $k_z$, absolute value of the transverse momentum, $\kappa$, and a definite TAM, $m$, consisting of the orbital part and helicity. The vector potential of a Bessel photon can be represented as a superposition of plane waves, whose momenta lie on the surface of a cone with an opening angle $\tan\theta_k = \kappa / k_z$
\begin{equation}
\label{vf}
    \bm{A}_{\kappa m k_z\Lambda}(\bm{r}) = \int a_{\kappa m k_z}(\bm{q})\bm{e}_{\bm{k}\Lambda} e^{i\bm{q}\cdot\bm{r} - i
    \omega t}\frac{d^3q}{(2\pi)^3}, \;\; a_{\kappa m k_z}(\bm{q}) = \frac{(2\pi)^2}{\kappa}\delta(q_{\perp} - 
    \kappa)\delta(q_z - k_z)e^{im\varphi_q},
\end{equation}
where $\bm{e}_{\bm{k}\Lambda}$ is the polarization vector of a plane wave with momentum $\bm{k}$, which in a Coulomb gauge satisifes
$\bm{e}_{\bm{k}\Lambda}\cdot\bm{k} = 0, \;\; \bm{e}_{\bm{k},\Lambda}\cdot\bm{e}^{*}_{\bm{k},
\Lambda'} = \delta_{\Lambda,\Lambda'}$. The explicit expression for the vector potential in a Coulomb gauge reads
\begin{equation}
\label{tw_p}
    \bm{A}_{\kappa m k_z \Lambda}(\bm{r}) = \sqrt{\frac{\pi \kappa }{3\omega LR}}e^{ik_z z - i\omega t}\sum\limits_{\sigma = 0, \pm 1 }d_{\sigma,\lambda}^{(1)}(\theta_k)e^{i(m-\sigma)\varphi_r}J_{m-\sigma}(\kappa \rho)\bm{\chi}_{\sigma}
\end{equation}

Here $d^{(1)}_{\lambda,\lambda'}(\theta_k)$ are the Wigner d-matrices \cite{Wigner1960, Varshalovich1988} and
\begin{equation}
    \bm{\chi}_{\pm} = \mp\frac{1}{\sqrt{2}}\begin{pmatrix}
        1\\\pm i\\0
    \end{pmatrix}, \bm{\chi}_0 = \begin{pmatrix} 0\\0\\1\end{pmatrix}
\end{equation} 
are the eigenvectors of the photon spin operator, corresponding to eigenvalues $\pm 1, 0 $ respectively. The vector potential is normalized as follows:
\begin{equation}
    \int \frac{\bm{E}^2 + \bm{H}^2}{8\pi}dV = \omega.
\end{equation}
The factor proportional to $1/\sqrt{LR}$ in \eqref{tw_p} arises since the vector potential is normalized in a large but finite volume. Thus, $L,R$ are the longitudinal and transverse scales of the quantization cylinder (the height and the radius of the cylinder base, respectively).

One of the quantum numbers of a twisted photon is $\Lambda = \pm 1$, often referred to as helicity. However, we should note that a twisted photon does not possess a definite value of momentum. Thus, the notion of helicity as of the photon spin projection onto the direction of its momentum loses meaningfulness. From \eqref{vf} it is clear that $\Lambda$ is the helicity of each plane wave in the decomposition of a twisted photon vector potential. However, different plane waves (with momenta $\bm{k}$ and $\bm{k}'$) in the decomposition propagate in different directions. From Eq.\eqref{tw_p} we see that the vector potential of a twisted photon is a superposition of eigenstates of the $z$ - projection spin operator 
\begin{equation}
    \hat{s}_z = \begin{pmatrix}
        0 & -i & 0\\
        i & 0 & 0\\
        0 & 0 & 0
    \end{pmatrix}.
\end{equation}
It possesses a spin projection onto the average propagation direction $\sigma = 0,\pm 1$ and OAM $m-\sigma$. The relative contribution of different spin-eigenstates with $\sigma$ and $\sigma'$ to the vector potential of a twisted photon is proportional to $d_{\sigma,\Lambda}^{(1)}(\theta_k)/d_{\sigma',\Lambda}^{(1)}(\theta_k)$. Therefore, for a given longitudinal momentum and energy of a twisted photon we can think of $\Lambda$ as a quantity that determines the  contribution of a spin-eigenstate to the vector potential. For example, for $\Lambda = 1$ it can be checked that $d_{1,\Lambda}^{(1)}(\theta_k)/d_{-1,\Lambda}^{(1)}(\theta_k) > 1$. This can be seen even more clearly if we consider the vector potential \eqref{tw_p} in the paraxial approximation ($k_z \gg \kappa)$

\begin{equation}
\label{par}
    \bm{A}_{\kappa m k_z\Lambda}(\bm{r}) \propto J_{m - \Lambda}(\kappa \rho)e^{i(m - \Lambda)\varphi + i k_z z}\bm{\chi}_{\Lambda},
\end{equation}
In \eqref{par} we can explicitly see that $\Lambda$ describes the photon spin and simply determines the direction of the vector potential rotation in the transverse plane (relative to its propagation axis). The sign of helicity determines the direction of rotation with $\Lambda = 1$ corresponding to counterclockwise rotation and $\Lambda = -1$ to clockwise rotation. Also, from the simple paraxial expression we can see that vortex photons with a given TAM $m$ but different helicities have different spatial profiles (and, hence, energy distributions) as they are proportional to $J_{m \mp 1}^2(\kappa \rho)$. In particular, the first ring in the spatial energy distribution of a photon with $\Lambda = -1$ will be located further from the photon propagation axis.

\section{Matrix element of vortex photon absorption}
\label{sec:mat_elem}
Let us now obtain the S-matrix element for the absorption of a vortex photon by an electron in a Landau state. In the first order of perturbation theory the matrix element reads:
\begin{equation}
\label{s}
    S^{(1)}_{fi} = \langle e_f | \hat{S} | e_i,\gamma\rangle = -ie \int d^4x j^{\mu}_{fi}(x)A_{\mu}(x).
\end{equation}
Here, $|e_i,\gamma\rangle = |e_i\rangle\otimes|\gamma\rangle$ is the vector describing the state of the initial electron and the incident photon, $|e_f \rangle$ is the final state of the electron after the photon absorption, $A^{\mu} = (0,\bm{A})$ is the incident photon vector potential \eqref{tw_p}. The transition current is given by:
\begin{equation}
\label{j}
    j^{\mu}_{fi}(x) = \overline{\Psi}_f(x)\gamma^{\mu}\Psi_i(x),
\end{equation}
with $\Psi_{i,f}(x)$ being the corresponding wave functions of the initial and the final electron states, respectively, in the coordinate representation. The explicit expressions for the transition current are given in Appendix \ref{sec:app_4cur}.

For an electron transition between the Landau states with quantum numbers $s_i, j_i$ and $s_f, j_f$ for the ``spin-up'' initial and final electrons the $S$-matrix element is
\begin{multline} \label{eq:S_up_up}
     S_{\uparrow\uparrow}^{(1)} = 
     \sqrt{\frac{2\pi\kappa}{3\omega LR}} ie N_i N_f(2\pi)^3 \rho_H \delta(\varepsilon_i - \varepsilon_f + \omega)\delta(p_{zi} - p_{zf} + k_z)\delta_{j_f,j_i+m} \times \\ \times \Bigg[2\sqrt{2}\left((\varepsilon_i + m_e)d^{(1)}_{-1\Lambda}(\theta_k)\mathcal{F}_{s_f,s_i}^{j_f+1/2,j_i-1/2}(\kappa\rho_H) - (\varepsilon_f + m_e)d_{1\Lambda}^{(1)}(\theta_k)\mathcal{F}_{s_f,s_i}^{j_f-1/2,j_i+1/2}(\kappa\rho_H)\right) + \\
     + \rho_H d^{(1)}_{0\Lambda}(\theta_k) \left(m_e(p_{zi} + p_{zf}) + p_{zf}\varepsilon_i + p_{zi}\varepsilon_f\right)\mathcal{F}_{s_f,s_i}^{j_f-1/2,j_i-1/2}(\kappa\rho_H)\Bigg],
\end{multline}
where
\begin{equation}
\label{Ffunc}
    \mathcal{F}^{\ell,\ell'}_{s,s'}(y) =  \frac{(s'+\ell')!}{s!} \frac{1}{2^{2(s-s')+\ell-\ell'+1}}y^{2(s-s')+\ell-\ell'}L_{s'+\ell'}^{s-s'+\ell-\ell'}(y^2/4)L_{s'}^{s-s'}(y^2/4)e^{-y^2/4}.
\end{equation}
The function \eqref{Ffunc} can be found in \cite{GR1963}. Notice, that this edition contains a typo: $m \Leftrightarrow n$. The corrected expression \eqref{Ffunc} is also provided in \cite{Karlovets2023Sep} and \cite{Pavlov2024}. In \eqref{eq:S_up_up} $\varepsilon_i, p_{zi}$ and $\varepsilon_f, p_{zf}$ are the energy and the longitudinal momentum of the initial and the final electrons, respectively, $k_z$ and $\kappa$ refer to the longitudinal momentum and the absolute value of the transverse momentum of the absorbed photon, respectively. For the $S$-matrix elements with other configurations of the initial and the final electron spins one is referred to Appendix \ref{sec:app_S_matr}. 

Notice that the function $\mathcal{F}_{s,s'}^{\ell,\ell'}(y)$ and therefore the entire $S$-matrix includes an exponentially decreasing factor. The same expressions (with the only difference being the complex conjugation of $A^\mu$) are obtained for photon emission by an electron in a strong magnetic field \cite{Karlovets2023Sep, Pavlov2024}. This indicates that the probabilities of absorbing (and emitting) a vortex photon include a damping factor
\begin{equation}
\label{whyparax}
    |S_{fi}^{(1)}|^2 \propto \exp(-\frac{(\kappa/\kappa_c)^2}{2}), \;\; \kappa_c = \frac{\sqrt{2}}{\rho_H} = m_e\sqrt{\frac{H}{H_c}}
\end{equation}
and are larger for smaller absolute values of the photon transverse momentum, $\kappa \ll \kappa_c$. The critical value $\kappa_c$ is of the order $100\,\text{keV} - 1\,\text{MeV}$ for $H \sim (0.01 - 1)\, H_c$.

\section{Kinematic requirements}
\label{sec:kinem}
Let us now briefly discuss the kinematic properties of the absorption process. The $\delta$-functions and the Kronecker symbol in \eqref{eq:S_up_up} reflect the conservation laws:
\begin{equation} \label{eq:ener_con_law}
    \varepsilon_f= \varepsilon_i + \omega,
\end{equation}
\begin{equation} \label{eq:lp_con_law}
    p_{zf} = p_{zi} + k_z,
\end{equation}
\begin{equation} \label{eq:j_con_law}
    j_f = j_i + m.
\end{equation}
For a transition between the electron quantum states characterized by quantum numbers $s_i,j_i \rightarrow s_f,j_f$ with a given energy and longitudinal momentum of the incident photon $\omega, k_z$ we can solve the equations \eqref{eq:ener_con_law} - \eqref{eq:j_con_law} for the longitudinal momentum of the initial electron. We find it to be
\begin{equation} \label{eq:p+-}
    p_{zi}^{\pm} = \frac{1}{2\kappa^2}\left[k_z\zeta_{s_f s_i m} \pm \omega\sqrt{\zeta_{s_f s_i m}^2 - 4\kappa^2\left[m_e^2 + \frac{4}{\rho_H^2}\left(s_i + j_i + \frac{1}{2}\right)\right]}\right],
\end{equation}
where 
\begin{equation} \label{eq:zeta}
    \zeta_{s_f,s_i,m} = 4\rho_H^{-2}(s_f-s_i+m) - \kappa^2.
\end{equation} 
There are two possible solutions, such that $p_{zi}^{+} > k_z$ and $p_{zi}^{-} < k_z$. From this simple consideration it follows that absorption cross sections as functions of the longitudinal momentum of the initial electron should represent a discrete spectrum. This is reasonable because the transverse energy of the electron states is quantized and the conservation laws \eqref{eq:ener_con_law} can not be satisfied for an arbitrary electron longitudinal momentum. However, this is not the case for the photon emission problems, where the photon energy is not a fixed parameter, but is determined by the transition energy $\varepsilon_f - \varepsilon_i$. Alternatively, one could consider the initial momentum of the electron to be fixed and obtain a restriction on the photon energy that would allow absorption accompanied by the electron transition $s_i,j_i \rightarrow s_f,j_f$. In the reference frame corresponding to $p_{zi} = 0$ (from hereon referred to as the electron ``rest frame''), we find the photon energy required for absorption to be
\begin{equation}
\label{eq:kinematic}
    \omega^{(a)} = (s_f - s_i + m) \omega_c - \frac{\kappa^2}{2\varepsilon_i} = \frac{2(s_f - s_i + m)\omega_c}{\sqrt{1 + \displaystyle \frac{2\omega_c\sin^2\theta_k}{\varepsilon_i}(s_f - s_i + m)} + 1},
\end{equation}
where $\omega_c = eH/\varepsilon_i$ is the cyclotron frequency and $\theta_k$ is the conical angle of the incident photon. We can see that $\omega^{(a)}$ are not exact multiples of the cyclotron frequency due to the electron recoil caused by the photon non-zero transverse momentum (see, e.g., \cite{Harding1991Jun}). However, as the incident photon becomes more delocalized, the recoil decreases. In the limit $\theta_k \rightarrow 0$, the frequency $\omega^{(a)}$ approaches $N\omega_c$, where $N \in \mathbb{N}$.

\section{Results}
\label{sec:results}
We aim to analyze absorption cross sections for various parameters such as the TAM of the initial electron, $j_i$, the TAM of the photon, $m$, the transverse photon momentum, $\kappa$, and others. It should be noted that the cross section here is generalized to the states of the initial electron and the incident photon possessing no fixed value of their momenta (see Sec. \ref{app:cross_sect} in Appendix for detailed information). We will focus on two specific quantities. The first one is the absorption cross section summed over the spin states of the final electron, $\sigma_z$, and averaged over the incident photon helicity, $\Lambda$, 
\begin{equation}
\label{sig1}
    \sigma_{\text{tot}} = \frac{1}{2}\sum\limits_{\Lambda = \pm 1 }\sum\limits_{\sigma_{zf} = \pm 1/2} \int d\sigma.
\end{equation}
The second one is the absorption cross section only averaged over the incident photon helicity,
 \begin{equation}
 \label{sig2}
     \sigma_{\text{spin}} = \frac{1}{2}\sum\limits_{\Lambda = \pm 1} \int d\sigma.
 \end{equation}
Here $d\sigma$ is defined by \eqref{eq:diff_sig}. Notice, that in a set up with the given longitudinal momentum of the initial electron and the given energy and longitudinal momentum of the incident photon the quantum numbers of the final electron, $s_f,j_f,p_{zf}$, are fixed by the conservation laws \eqref{eq:ener_con_law}-\eqref{eq:j_con_law}. Therefore, $\sigma_{\text{tot}}$ in Eq.\eqref{sig1} represents total absorption cross section. Alternatively, the quantity $\sigma_{\text{spin}}$ \eqref{sig2} allows us to compare absorption processes involving different spin transitions and analyze their contribution to total cross section. In this work we primarily focus on how TAM of the particles affects absorption processes and are not interested in polarization related effects. On top of that polarization resolved measurements are often challenging not only in astrophysical observations, but also in terrestrial experiments. Therefore, we take averages over $\Lambda$ in Eq.\eqref{sig1},\eqref{sig2}.

Since the cross section is a Lorentz invariant quantity \cite{Kotkin1992, Karlovets2020Apr}, it is convenient to analyze it in the ``rest frame'' of the initial electron. The energy of the incident photon is then determined by the conservation laws and changes for various parameters such as the TAM of the initial electron and the incident photon. For the parameters we use below the photon energy lies in the range $10 \,\text{keV} - 10 \,\text{MeV}$, which corresponds to hard X-rays and gamma rays.

\begin{figure}
    \centering
    \includegraphics[width=1\linewidth]{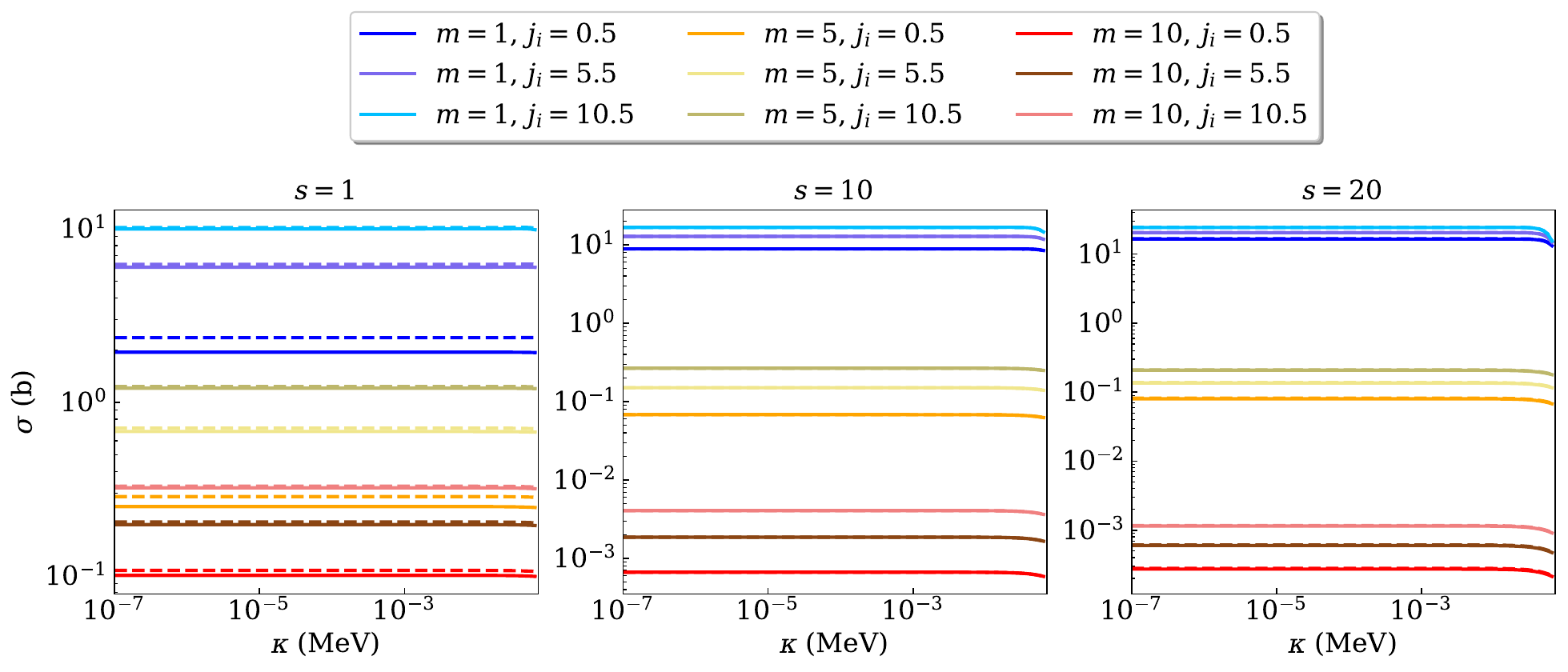}
    \caption{Total absorption cross sections $\sigma_{\text{tot}}$ as functions of the transverse momentum $\kappa$ of the photon. $H = H_c$. Left panel corresponds to radial quantum numbers of initial and final electron $s_i = s_f = 1$, middle - $s_i = s_f = 10$, right panel - $s_i = s_f = 20$. Solid lines correspond to ``spin-up'' initial electron state, dashed lines - to ``spin-down'' initial electron state.}
    \label{fig:sig_kappa_s=1_20}
\end{figure}

In Fig. \ref{fig:sig_kappa_s=1_20} we investigate the dependence of total absorption cross section $\sigma_{\text{tot}}$ on the transverse momentum of the incident photon. From hereon, cross sections are expressed in barns ($1\,\text{b} = 10^{-24} \,\text{cm}^{-2}$). As discussed earlier in Sec. \ref{sec:mat_elem} the matrix elements inclide an exponential damping factor $\exp( - \kappa^2 / (2\kappa_{\text{c}}^2))$. Therefore, the emission and the absorption of photons with $\kappa \gtrsim \kappa_{\text{c}}$ (0.511 MeV for $H \sim H_{\text{c}}$) is suppressed. For smaller values of $\kappa$ absorption cross sections do not depend on $\kappa$. Note that although cross sections in Fig. \ref{fig:sig_kappa_s=1_20} are given only for $H = H_c$ their behaviour is also the same for $H < H_c$. Cross sections are also influenced by the other parameters, increasing when TAM of the incident photon is lower and the TAM of the initial electron is higher. For large radial quantum numbers ($s \ge 20$) the cross sections for the same photon TAM group together with the separation between these groups growing as $s$ increases.

\begin{figure}[h!]
    \centering
    \includegraphics[width=1\linewidth]{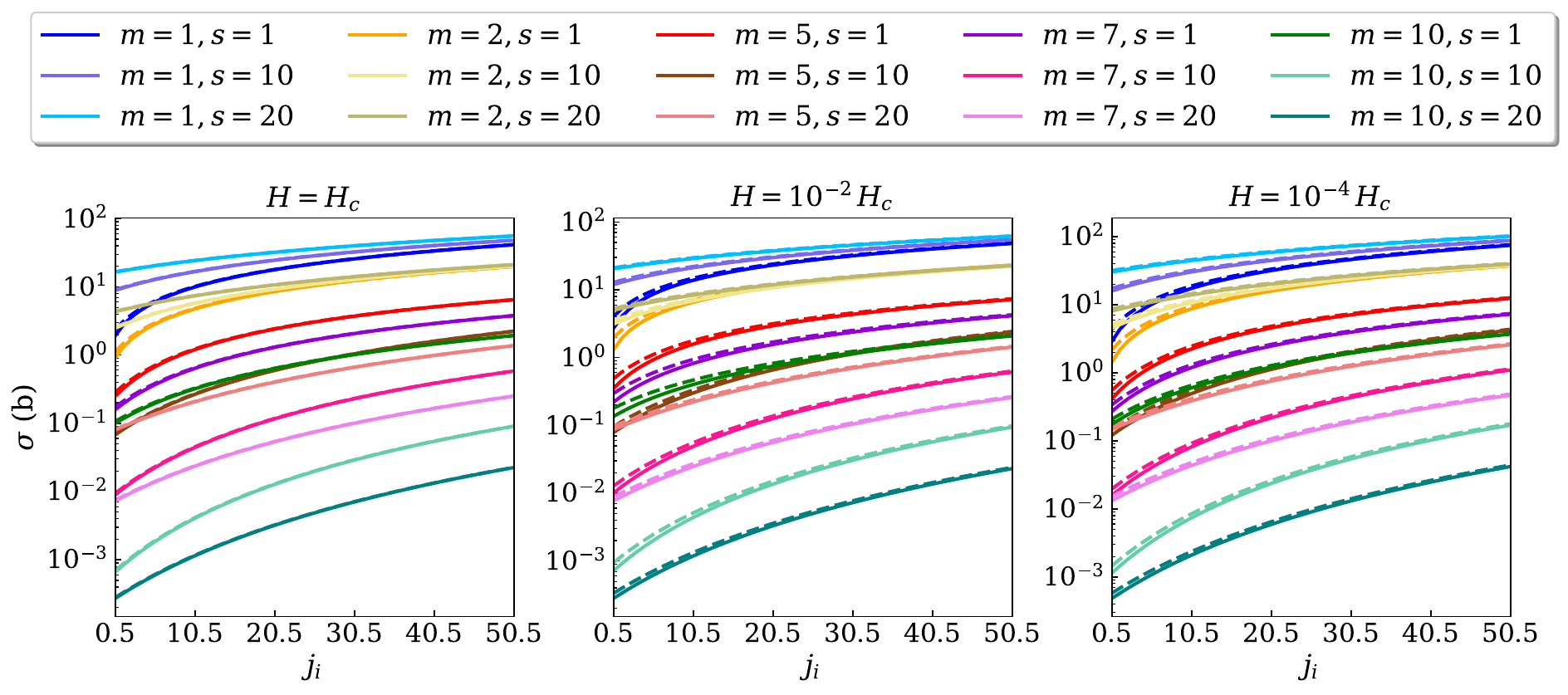}
    \caption{Total absorption cross sections \eqref{sig1} of a photon with $\kappa = 10$ eV as functions of the initial electron TAM $j_i$ for different TAM of the incident photon $m$ and electron radial quantum numbers $s_i = s_f$. Different colors represent different TAM of the photon. The solid and dashed lines stand for the initial electron spin state (``spin-up'' and ``spin-down'', respectively)}
    \label{fig:sig_H=1}
\end{figure}

Next we examine how $\sigma_{\text{tot}}$ changes with the initial electron TAM $j_i$. We focus on cases when the radial quantum numbers of the initial and the final electron states are the same, as transitions where $s_i = s_f$ are the most dominant. It is straightforward to check, that for $s_i \ne s_f$ absorption probabilities decrease due to a smaller overlap of the initial and final electron wave functions. However, it is important to note that if the energy, momentum and other quantum numbers of the incident photon and initial electron are fixed, the radial quantum number of the final electron state is determined by the energy-momentum conservation laws. In Fig.\ref{fig:sig_H=1} we see that cross sections steadily increase with increasing values of $j_i$. Consistent with previous observation the cross sections are smaller for higher photon TAM. The solid and dashed lines in Fig.\ref{fig:sig_H=1} represent different spin states of the initial electron. While there is almost no difference for them in a critical magnetic field, it becomes more pronounced as the magnetic field decreases.

Next we focus on spin resolved transitions by analyzing $\sigma_{\text{spin}}$. In Fig.\ref{fig:sflips_s=1} we compare absorption cross sections for different electron spin transitions, various TAM of the incoming photon and different radial quantum numbers, all as functions of $j_i$. We observe that the absorption without a spin flip dominates, and the cross sections for the ``spin up - spin up'' and ``spin down'' - ``spin down'' electron transitions are much closer to each other compared to those involving the spin flip. Additionally, ``spin down'' - ``spin up'' transitions are more prominent than ``spin up'' - ``spin down'' ones. As $j_i$ increases, the cross sections for non-spin-flip processes grow, while those for spin-flip processes slightly decrease. It is clear that the growth of the cross sections for non-spin-flip processes is greater than the decrease of those with a spin flip as the total absorption cross section grows with $j_i$. 

The most interesting feature in Fig.\ref{fig:sflips_s=1} is the presence of a sharp collapse of cross sections at certain value of $j_i$. In Fig.\ref{fig:sflips_s=1} it is observed for the ``spin up'' - ``spin down'' electron transition corresponding to the incident photon TAM $m = 5$ (left top panel). We can roughly estimate when this occurs by setting the $S$-matrix elements to zero and solving it for the incident photon energy. There is no solution for processes without a spin flip. However, for spin-flip processes, assuming $\kappa \ll \omega, \kappa \rho_{\text{H}} \ll 1, s_i = s_f$ we can calculate the photon energies where this collapse happens:

\begin{figure}[t]
    \centering
    \includegraphics[width=1\linewidth]{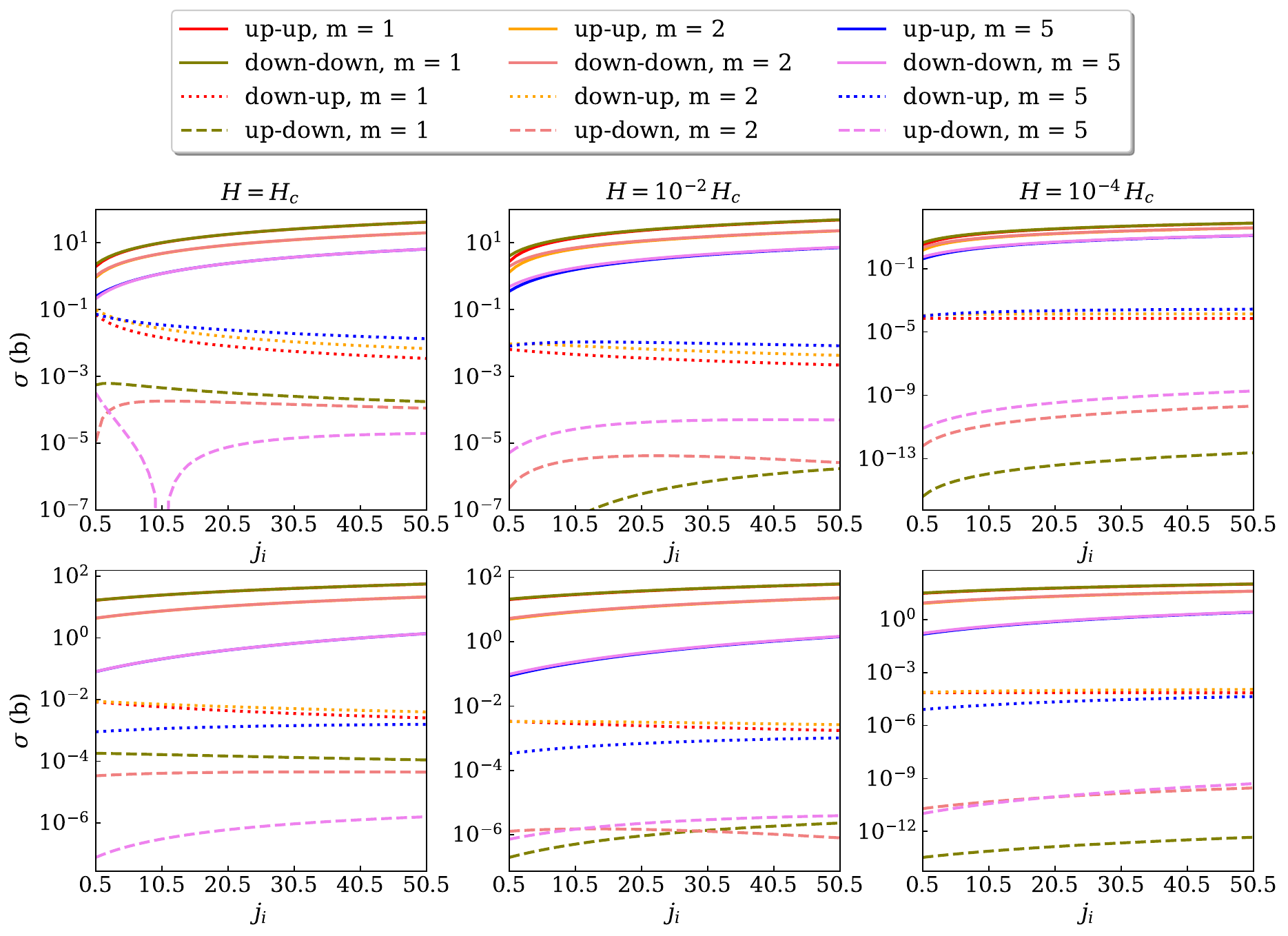}
    \caption{Absorption cross sections \eqref{sig2} for different spin states (up and down) of initial and final electron. $\kappa = 10$ eV. Upper row corresponds to $s_i = s_f = 1$, lower row - $s_i = s_f = 20$.}
    \label{fig:sflips_s=1}
\end{figure}

\begin{equation}
\label{ener_coll_occur}
    \omega_{\downarrow\uparrow,\rm{min}} \approx m_e\frac{H}{H_c} \frac{\Lambda}{\Lambda - 1}\frac{2(m + 1)}{(1 + \varepsilon/m_e)}, \quad \omega_{\uparrow\downarrow,\rm{min}} \approx \frac{\kappa^2}{m_e}\frac{\Lambda}{\Lambda + 1}\frac{(s_i + j_i + 1/2)}{m(1 + \varepsilon_i/m_e)}
\end{equation}
We highlight that this expressions are obtained for the photon energy in the initial electron ``rest frame''. The key observation is that these energies diverge depending on the incident photon helicity either for $\Lambda = 1$ or $\Lambda = -1$ depending on the specific electron spin transition. Thus, the sharp minima can only occur when a photon with $\Lambda = -1$ induces a ``spin-up'' $\rightarrow$ ``spin-down'' electron transition or when a photon with $\Lambda = 1$ triggers a ``spin-down'' $\rightarrow$ ``spin-up'' electron transition. The photon energies \eqref{ener_coll_occur} can be estimated as:
\begin{equation}
    \begin{aligned}
        & \omega_{\downarrow\uparrow,\text{min}} \sim m_e\left(H/H_{\rm{c}}\right) \sim 0.511 \,\text{MeV} \quad (H \sim H_c)\\ & \omega_{\uparrow\downarrow,\text{min}} \sim \kappa^2/m_e \sim 0.25\,\text{eV} \quad (\kappa = 0.5\,\text{keV}).
    \end{aligned}
\end{equation}    

As previously discussed, the considered photon energy range for the parameters correspond to hard X-rays and gamma rays (from tens of keV to several MeV). Therefore, the energy $\omega_{\uparrow\downarrow,\text{min}}$ falls outside this range and one can only observe the sharp minima for the ``spin-up'' $\rightarrow$ ``spin-down'' electron transition. To estimate TAM of the initial electron that corresponds to the absorption of a photon with the energy $\omega_{\downarrow\uparrow,\text{min}}$ we compare it with \eqref{eq:kinematic} and find the following:
\begin{equation}
\label{wheremin}
    j_{\rm{min}} \approx \frac{1}{2}\frac{H_c}{H}(m^2 - 1) - s - \frac{1}{2}.
\end{equation}
Note that the expression on the right hand side of \eqref{wheremin} is not required to be a half-integer, therefore the sharp minimum actually occurs at the nearest half-integer value to $j_\text{min}$. For example, in a critical magnetic field, incident photon TAM $m = 5$ and radial quantum numbers $s_{i} = s_{f} = 1$ (top left panel in Fig.\ref{fig:sflips_s=1}) we find $j_{\text{min}} = 10.5$, which matches the exact sharp minimum's position. Eq.\eqref{wheremin} also explains the absence of sharp minima in the bottom row in Fig.\ref{fig:sflips_s=1} where $s_{i} = s_{f} = 20$. For given values of $m$ and $H$, $j_{\text{min}}$ is either negative or greater than $50.5$. Therefore, the sharp minima falls outside the range shown in the figure.

\section{Conclusion and outlook}
\label{sec:disc_n_conc}

In this work we have addressed the problem of a twisted photon absorption by an electron in a strong magnetic field. The absorption cross sections have been studied and their dependence on the quantum numbers of the initial particles has been analyzed. It has been found that for a given TAM of the initial electron and incident photon absorption cross sections remain independent on the absolute value of the incident photon transverse momentum, $\kappa$, as it varies from $0.1$ eV to $100$ keV. However, for the larger values, $\kappa \gtrsim m_e\sqrt{H/H_c}$, absorption cross section is exponentially suppressed. On top of that cross sections corresponding to larger radial quantum numbers and the same incident photon TAM begin to group into distinct bunches.

Next, it has been shown that total absorption cross sections grow for larger values of the TAM of the initial electron and decrease for larger values of the TAM of the incident photon. In the critical magnetic field ($H \sim H_c$) the cross section of the absorption of a photon with $m = 1$ exceeds the Thompson scattering cross section ($\sigma_T \sim 1$ b) by $1-2$ orders. It has also been noted that as the magnetic field strength increases, the difference between the cross sections for opposite spin states of the initial electron disappears, though the overall values of the cross sections remain largely unchanged.

Next, we have analyzed the absorption of a photon for given spin states of the initial and final electrons and compared the cross sections for different spin transitions. First, the cross sections for processes where the electron’s spin remains the same (no spin flip) are several orders of magnitude higher than those where the electron’s spin changes. On top of that, there is much less difference in the absorption cross sections without electron spin transition (``spin-up'' $\rightarrow$ ``spin-up'' and ``spin-down'' $\rightarrow$ ``spin-down'') compared to those for the absorption with a spin flip (``spin-down'' $\rightarrow$ ``spin-up'' and ``spin-up'' $\rightarrow$ ``spin-down''). It has been shown that spin-flip cross sections are always larger for the ``spin-down'' $\rightarrow $ ``spin-up'' transition, which could be interpreted as the analogy of the Sokolov-Ternov effect present in photon emission \cite{ST1974, vanKruining2019Sep}. It has also been found that the cross section for the reverse ``spin-up'' $\rightarrow$ ``spin-down'' transition sharply decreases at certain values of the initial electron TAM.

We believe that studying the absorption of twisted photons by electrons in a strong magnetic field could help improve our understanding of radiative transport in astrophysical environments. Unlike plane-wave photons, twisted photons can excite electrons to states with higher angular momentum. The emission and absorption of twisted photons by electrons could play a key role in processes occurring in neutron star magnetospheres, kilonovae, jets of active galactic nuclei and others. Additionally, accounting for these processes could help explain certain features in the spectra and light curves of objects with pulsed emission, such as X-ray transients and soft gamma repeaters (see, e.g., \cite{Boldt1976Jul, Lutovinov2015Apr, Thompson1995Jul, Lyubarsky2002May}).

There could also be the several following extensions to this research. First, one could look into the absorption of a photon approaching an electron at a specific impact parameter and further investigate an absorption by a localized ensemble of electrons described by a specific distribution function. Moreover, it could be of interest to study absorption cross sections in a different gauge of the magnetic field vector potential, namely, the Hermite-Gaussian solutions. Also, notice that the absorption of the photon by an electron in a magnetic field is only the specific (resonant) case of the broader process of the electron-photon scattering (Compton effect). Therefore, the analogous treatment for the Compton effect stimulated by a magnetic field and accounting for a photon vorticity also takes place for the future works.

\section{Acknowledgments}
We are grateful to I. Pavlov, A. Chaikovskaia, G. Sizykh, D. Glazov, and N. Sheremet for useful discussions and criticism. The studies in Secs. II-III are supported by the Russian Science Foundation (Project \textnumero 23-62-10026, https://rscf.ru/en/project/23-62-10026/). The work in the Secs. IV-V on the derivation of absorption matrix elements (by D. Karlovets) is supported by the Foundation for the Advancement of Theoretical Physics and Mathematics “BASIS”.
The studies in Sec. VI is supported by the Ministry of Science and Higher Education of the Russian Federation (agreement \textnumero 075-15-2021-1349). The work in Sec.VII on the analysis of the obtained results is supported by the Government of the Russian Federation through the ITMO Fellowship and Professorship Program. 

\appendix
\section{Appendix: Transition currents}
\label{sec:app_4cur}

The transition current components for a twisted photon absorption and the ``spin-up'' $\rightarrow$ ``spin-up'' electron transition are
\begin{multline}
    j_{\uparrow\uparrow}^0(x) = N_i^\uparrow N_f^\uparrow e^{i((p_{zi} - p_{zf})z + (j_i - j_f)\varphi_r - (\varepsilon_i - \varepsilon_f)t)}\tilde\rho^{j_i + j_f - 1}e^{-\tilde\rho^2}\Bigg((m_e^2 + p_{zi} p_{zf} + \varepsilon_i\varepsilon_f + m_e(\varepsilon_i + \varepsilon_f)) L_{s_i}^{j_i - 1/2}(\tilde\rho^2)L_{s_f}^{j_f - 1/2}(\tilde\rho^2) + \\ \frac{4}{\rho_H^2}\tilde\rho^2L_{s_i}^{j_i + 1/2}(\tilde\rho^2)L_{s_f}^{j_f + 1/2}(\tilde\rho^2)\Bigg)
\end{multline}
\begin{multline}
    j_{\uparrow\uparrow}^1(x) = \frac{2i}{\rho_H}N_i^\uparrow N_f^\uparrow e^{i((p_{zi} - p_{zf})z + (j_i - j_f)\varphi_r - (\varepsilon_i - \varepsilon_f)t)} \tilde\rho^{j_i + j_f} e^{-\tilde\rho^2} \times \\ \times \Big((\varepsilon_f + m_e)e^{i\varphi_r}L_{s_f}^{j_f - 1/2}(\tilde\rho^2)L_{s_i}^{j_i+1/2}(\tilde\rho) - (\varepsilon_i + m_e)e^{-i\varphi_r}L_{s_f}^{j_f+1/2}(\tilde\rho^2)L_{s_i}^{j_i - 1/2}(\tilde\rho^2)\Big)
\end{multline}

\begin{multline}
   j_{\uparrow\uparrow}^2(x) = \frac{2}{\rho_H}N_i^\uparrow N_f^\uparrow e^{i((p_{zi} - p_{zf})z + (j_i - j_f)\varphi_r - (\varepsilon_i - \varepsilon_f)t)} \tilde\rho^{j_i + j_f} e^{-\tilde\rho^2} \times \\ \times \Big((\varepsilon_f + m_e)e^{i\varphi_r}L_{s_f}^{j_f-1/2}(\tilde\rho^2)L_{s_i}^{j_i+1/2}(\tilde\rho) + (\varepsilon_i + m_e)e^{-i\varphi_r}L_{s_f}^{j_f+1/2}(\tilde\rho^2)L_{s_i}^{j_i-1/2}(\tilde\rho^2)\Big)
\end{multline}

\begin{equation}
   j_{\uparrow\uparrow}^3(x) = N_i^\uparrow N_f^\uparrow e^{i((p_{zi} - p_{zf})z + (j_i - j_f)\varphi_r - (\varepsilon_i - \varepsilon_f)t)}e^{-\tilde\rho^2}\Big(m_e(p_{zi} + p_{zf}) + p_{zf}\varepsilon_i + p_{zi}\varepsilon_f\Big)\tilde\rho^{j_i + j_f - 1}L_{s_f}^{j_f-1/2}(\tilde\rho^2)L_{s_i}^{j_i-1/2}(\tilde\rho).
\end{equation}
The indices $i$ and $f$ stand for the initial and final electron spin states respectively. Note, that the first index of $j(x)$ designates the spin state of the \textit{final} electron, while the second one is for the \textit{initial} electron according to \eqref{j}. For the ``spin-up'' -- ``spin-down'' transition we have:
\begin{multline}
    j_{\downarrow\uparrow}^0(x) = \frac{2i}{\rho_H} N_i^\uparrow N_f^\downarrow e^{i((p_{zi} - p_{zf})z + (j_i - j_f)\varphi_r - (\varepsilon_i - \varepsilon_f)t)}\tilde\rho^{j_i + j_f - 1} e^{-\tilde\rho^2}\times \\ \times \left[p_{zi}\left(s_f + j_f + \frac{1}{2}\right)L_{s_i}^{j_i-1/2}(\tilde\rho^2) L_{s_f}^{j_f-1/2}(\tilde\rho^2) - p_{zf} \tilde\rho^2 L_{s_i}^{j_i+1/2}(\tilde\rho^2) L_{s_f}^{j_f+1/2}(\tilde\rho^2)\right]
\end{multline}

\begin{equation}
    j_{\downarrow\uparrow}^1(x) = -N_i^\uparrow N_f^\downarrow e^{i((p_{zi} - p_{zf})z + (j_i - j_f - 1)\varphi_r - (\varepsilon_i - \varepsilon_f)t)}e^{-\tilde\rho^2}\Big(m_e(p_{zf} - p_{zi}) + p_{zf}\varepsilon_i - p_{zi}\varepsilon_f\Big)\tilde\rho^{j_i + j_f}L_{s_i}^{j_i-1/2}(\tilde\rho^2)L_{s_f}^{j_f+1/2}(\tilde\rho^2)
\end{equation}

\begin{equation}
    j_{\downarrow\uparrow}^2(x) = -iN_i^\uparrow N_f^\downarrow e^{i((p_{zi} - p_{zf})z + (j_i - j_f - 1)\varphi_r - (\varepsilon_i - \varepsilon_f)t)}e^{-\tilde\rho^2}\Big(m_e(p_{zf} - p_{zi}) + p_{zf}\varepsilon_i - p_{zi}\varepsilon_f\Big)\tilde\rho^{j_i + j_f}L_{s_i}^{j_i-1/2}(\tilde\rho^2)L_{s_f}^{j_f+1/2}(\tilde\rho^2)
\end{equation}

\begin{multline}
   j_{\downarrow\uparrow}^3(x) = \frac{2i}{\rho_H}N_i^\uparrow N_f^\downarrow e^{i((p_{zi} - p_{zf})z + (j_i - j_f)\varphi_r - (\varepsilon_i - \varepsilon_f)t)} \tilde\rho^{j_i + j_f - 1}e^{-\tilde\rho^2} \times \\ \times
   \left[\left(s_f + j_f + \frac{1}{2}\right)(\varepsilon_i + m_e)L_{s_i}^{j_i - 1/2}(\tilde\rho^2)L_{s_f}^{j_f-1/2}(\tilde\rho^2) - (\varepsilon_f + m_e) L_{s_i}^{j_i + 1/2}(\tilde\rho^2)L_{s_f}^{j_f + 1/2}(\tilde\rho^2)\right].
\end{multline}
Analogously, for the ``spin-down'' $\rightarrow$ ``spin-up'' transition:
\begin{multline}
    j_{\uparrow\downarrow}^0(x) = -\frac{2i}{\rho_H}N_i^\downarrow N_f^\uparrow e^{i((p_{zi} - p_{zf})z + (j_i - j_f)\varphi_r - (\varepsilon_i - \varepsilon_f)t)} \tilde\rho^{j_i + j_f - 1}e^{-\tilde\rho^2} \times \\
    \times \left[p_{zf}\left(s_i + j_i + \frac{1}{2}\right) L_{s_i}^{j_i - 1/2}(\tilde\rho^2) L_{s_f}^{j_f - 1/2}(\tilde\rho^2) - p_{zi}\tilde\rho^2 L_{s_i}^{j_i + 1/2}(\tilde\rho^2) L_{s_f}^{j_f + 1/2}(\tilde\rho^2)\right]
\end{multline}
\begin{equation}
    j_{\uparrow\downarrow}^1(x) = N_i^\downarrow N_f^\uparrow e^{i((p_{zi} - p_{zf})z + (j_i - j_f + 1)\varphi_r - (\varepsilon_i - \varepsilon_f)t)}e^{-\tilde\rho^2}(m_e(p_{zf} - p_{zi}) + p_{zf}\varepsilon_i - p_{zi}\varepsilon_f)\tilde\rho^{j_i + j_f} L_{s_i}^{j_i + 1/2}(\tilde\rho^2)L_{s_f}^{j_f - 1/2}(\tilde\rho^2)
\end{equation}
\begin{equation}
    j_{\uparrow\downarrow}^2(x) = -i N_i^\downarrow N_f^\uparrow e^{i((p_{zi} - p_{zf})z + (j_i - j_f + 1)\varphi_r - (\varepsilon_i - \varepsilon_f)t)}e^{-\tilde\rho^2}(m_e(p_{zf} - p_{zi}) + p_{zf}\varepsilon_i - p_{zi}\varepsilon_f)\tilde\rho^{j_i + j_f} L_{s_i}^{j_i + 1/2}(\tilde\rho^2)L_{s_f}^{j_f - 1/2}(\tilde\rho^2)
\end{equation}
\begin{multline}
    j_{\uparrow\downarrow}^3(x) = -\frac{2i}{\rho_H} N_i^\downarrow N_f^\uparrow e^{i((p_{zi} - p_{zf})z + (j_i - j_f)\varphi_r - (\varepsilon_i - \varepsilon_f)t)}\tilde\rho^{j_i + j_f - 1} e^{-\tilde\rho^2} \times \\
    \times \left[\left(s_i + j_i + \frac{1}{2}\right)(m_e + \varepsilon_f) L_{s_i}^{j_i - 1/2}(\tilde\rho^2) L_{s_f}^{j_f - 1/2}(\tilde\rho^2) - (m_e + \varepsilon_i)\tilde\rho^2 L_{s_i}^{j_i + 1/2}(\tilde\rho^2) L_{s_f}^{j_f + 1/2}(\tilde\rho^2)\right]
\end{multline}
Finally, for the ``spin-down'' $\rightarrow$ ``spin-down'' transition:
\begin{multline}
    j_{\downarrow\downarrow}^0(x) = N_i^\downarrow N_f^\downarrow e^{i((p_{zi} - p_{zf})z + (j_i - j_f)\varphi_r - (\varepsilon_i - \varepsilon_f)t)} \tilde\rho^{j_i + j_f - 1}e^{-\tilde\rho^2}\times \\ \times \left[\frac{4}{\rho_H^2}\left(s_i + j_i + \frac{1}{2}\right)\left(s_f + j_f + \frac{1}{2}\right) L_{s_i}^{j_i - 1/2}(\tilde\rho^2)L_{s_f}^{j_f - 1/2}(\tilde\rho^2) + (p_{zi} p_{zf} + (m_e + \varepsilon_i)(m_e + \varepsilon_f)) L_{s_i}^{j_i + 1/2}(\tilde\rho^2)L_{s_f}^{j_f + 1/2}(\tilde\rho^2)\right]
\end{multline}
\begin{multline}
    j_{\downarrow\downarrow}^1(x) = \frac{2i}{\rho_H} N_i^\downarrow N_f^\downarrow e^{i((p_{zi} - p_{zf})z + (j_i - j_f)\varphi_r - (\varepsilon_i - \varepsilon_f)t)}e^{-\tilde\rho^2}\tilde\rho^{j_i + j_f}\times \\ \times \left[e^{i\varphi_r}\left(s_f + j_f + \frac{1}{2}\right)(m_e + \varepsilon_i)L_{s_i}^{j_i + 1/2}(\tilde\rho^2) L_{s_f}^{j_f - 1/2}(\tilde\rho^2) - e^{-i\varphi_r}\left(s_i + j_i + \frac{1}{2}\right)(m_e + \varepsilon_f)L_{s_i}^{j_i - 1/2}(\tilde\rho^2)L_{s_f}^{j_f + 1/2}(\tilde\rho^2)\right]
\end{multline}
\begin{multline}
    j_{\downarrow\downarrow}^2(x) = \frac{2}{\rho_H} N_i^\downarrow N_f^\downarrow e^{i((p_{zi} - p_{zf})z + (j_i - j_f)\varphi_r - (\varepsilon_i - \varepsilon_f)t)}e^{-\tilde\rho^2}\tilde\rho^{j_i + j_f}\times \\ \times \left[e^{i\varphi_r}\left(s_f + j_f + \frac{1}{2}\right)(m_e + \varepsilon_i)L_{s_i}^{j_i + 1/2}(\tilde\rho^2) L_{s_f}^{j_f - 1/2}(\tilde\rho^2) + e^{-i\varphi_r}\left(s_i + j_i + \frac{1}{2}\right)(m_e + \varepsilon_f)L_{s_i}^{j_i - 1/2}(\tilde\rho^2)L_{s_f}^{j_f + 1/2}(\tilde\rho^2)\right]
\end{multline}
\begin{equation}
    j_{\downarrow\downarrow}^3(x) = N_i^\downarrow N_f^\downarrow e^{i((p_{zi} - p_{zf})z + (j_i - j_f)\varphi_r - (\varepsilon_i - \varepsilon_f)t)}e^{-\tilde\rho^2}(m_e(p_{zf} + p_{zi}) + p_{zf}\varepsilon_i + p_{zi}\varepsilon_f)\tilde\rho^{j_i + j_f + 1}L_{s_i}^{j_f + 1/2}(\tilde\rho^2)L_{s_f}^{j_f + 1/2}(\tilde\rho^2).
\end{equation}

\section{Appendix: Amplitudes and absorption probabilities}
\label{sec:app_S_matr}
To calculate the $S$-matrix element \eqref{s} we use the function \eqref{Ffunc} taken from \cite{GR1963}. We mention, that in this edition there is a typo in this formula. The corrected expression is
\begin{multline} \label{eq:Ffunc}
    \mathcal{F}^{\ell,\ell'}_{s,s'}(y) = \int_0^\infty dx x^{\ell+\ell'+1} L_s^{\ell}(x^2)L^{\ell'}_{s'}(x^2)J_{\ell-\ell'}(yx)e^{-x^2} = \\ \frac{(s'+\ell')!}{s!} \frac{1}{2^{2(s-s')+\ell-\ell'+1}}y^{2(s-s')+\ell-\ell'}L_{s'+\ell'}^{s-s'+\ell-\ell'}(y^2/4)L_{s'}^{s-s'}(y^2/4)e^{-y^2/4},
\end{multline}
which is also given correctly in \cite{Pavlov2024} and in \cite{Karlovets2023Sep}. For the first order $S$-matrix elements of twisted photon absorption by an electron in a Landau state obtained the following expressions for different spin projections $\sigma_z$ of the initial and final electron. For ``spin-up'' $\rightarrow$ ``spin-up'' transition we have:
\begin{multline}
     S_{\uparrow\uparrow}^{(1)} = 
     \sqrt{\frac{2\pi\kappa}{3\omega LR}} ie N_i^\uparrow N_f^\uparrow (2\pi)^3 \rho_H \delta(\varepsilon_i - \varepsilon_f + \omega)\delta(p_{zi} - p_{zf} + k_z)\delta_{j_f,j_i + m} \times \\ \Bigg[2\sqrt{2}\left((\varepsilon_i + m_e)d^{(1)}_{-1\Lambda}(\theta_k)\mathcal{F}_{s_f,s_i}^{j_f + 1/2,j_i - 1/2}(\kappa\rho_H) - (\varepsilon_f + m_e)d_{1\Lambda}^{(1)}(\theta_k)\mathcal{F}_{s_f,s_i}^{j_f - 1/2,j_i + 1/2}(\kappa\rho_H)\right) \\
     + \rho_H d^{(1)}_{0\Lambda}(\theta_k) \left(m_e(p_{zi} + p_{zf}) + p_{zf}\varepsilon_i + p_{zi}\varepsilon_f\right)\mathcal{F}_{s_f,s_i}^{j_f - 1/2,j_i - 1/2}(\kappa\rho_H)\Bigg].
\end{multline}
For the ``spin-up'' $\rightarrow$ ``spin-down'' transition:
\begin{multline}
    S_{\downarrow\uparrow}^{(1)} = -\sqrt{\frac{2\pi\kappa}{3\omega LR}} e(2\pi)^3 N_i^\uparrow N_f^\downarrow \rho_H \delta(\varepsilon_i - \varepsilon_f + \omega)\delta(p_{zi} - p_{zf} + k_z)\delta_{j_f,j_i + m} \\ \Bigg[\sqrt{2}\rho_H \left(m_e(p_{zi} - p_{zf}) - p_{zf}\varepsilon_i + p_{zi}\varepsilon_f\right)d_{-1\Lambda}^{(1)}(\theta_k)\mathcal{F}_{s_f,s_i}^{j_f + 1/2,j_i - 1/2}(\kappa\rho_H) + \\ 2 d_{0\Lambda}^{(1)}(\theta_k)\left((\varepsilon_i + m_e)(s_f + j_f + 1/2)\mathcal{F}_{s_f,s_i}^{j_f - 1/2,j_i - 1/2}(\kappa\rho_H) - (\varepsilon_f + m_e)\mathcal{F}_{s_f,s_i}^{j_f + 1/2,j_i + 1/2}(\kappa\rho_H)\right)\Bigg].
\end{multline}
For the ``spin-down'' $\rightarrow$ ``spin-up'' transition:
\begin{multline}
    S_{\uparrow\downarrow}^{(1)} = \sqrt{\frac{2\pi\kappa}{3\omega LR}} e\rho_H N_i^\uparrow N_f^\downarrow (2\pi)^3\delta(\varepsilon_i - \varepsilon_f + \omega)\delta(p_{zi} - p_{zf} + k_z)\delta_{j_f,j_i + m} \\ \Bigg[\sqrt{2}\rho_H \left(m_e(p_{zi} - p_{zf}) - p_{zf}\varepsilon_i + p_{zi}\varepsilon_f\right)d_{1\Lambda}^{(1)}(\theta_k)\mathcal{F}_{s_f,s_i}^{j_f - 1/2, j_i + 1/2}(\kappa\rho_H) + \\ 2d_{0\Lambda}^{(1)}(\theta_k)\left((s_i + j_i + 1/2)(m_e + \varepsilon_f) \mathcal{F}_{s_f,s_i}^{j_f - 1/2,j_i - 1/2}(\kappa\rho_H) - (m_e + \varepsilon) \mathcal{F}_{s_f,s_i}^{j_f + 1/2,j_i + 1/2}(\kappa\rho_H)\right)\Bigg].
\end{multline}
For the ``spin-down'' $\rightarrow$ ``spin-down'' transition:
\begin{multline}
    S_{\downarrow\downarrow}^{(1)} = \sqrt{\frac{2\pi\kappa}{3\omega LR}} ie\rho_H N_i^\downarrow N_f^\downarrow (2\pi)^3\delta(\varepsilon_i - \varepsilon_f + \omega)\delta(p_{zi} - p_{zf} + k_z)\delta_{j_f,j_i + m}\Bigg[2\sqrt{2}(s_i + j_i + 1/2)(m_e + \varepsilon_f)d_{-1\Lambda}^{(1)}(\theta_k)\mathcal{F}_{s_f,s_i}^{j_f + 1/2,j_i - 1/2}(\kappa\rho_H) - \\ (s_f + j_f + 1/2)(m_e + \varepsilon_i)d_{1\Lambda}^{(1)}(\theta_k)\mathcal{F}_{s_f,s_i}^{j_f - 1/2,j_i + 1/2}(\kappa\rho_H) + \rho_H d_{0\Lambda}^{(1)}(\theta_k)\left(m_e(p_{zi} + p_{zf}) + p_{zf}\varepsilon_i + p_{zi}\varepsilon_f\right)\mathcal{F}_{s_f,s_i}^{j_f + 1/2,j_i + 1/2}(\kappa\rho_H)\Bigg],
\end{multline}

The absorption probability for the given quantum numbers of the initial and final electrons and incident photon is related to the $S$-matrix element as
\begin{equation} \label{eq:diff_prob}
    dW = \left|S_{fi}^{(1)}\right|^2 \frac{L}{2\pi} dp_{zf}\Delta s_f \Delta j_f,
\end{equation}
where $\Delta s_f = \Delta j_f = 1$. We use the next rule for squaring matrix elements in calculations: 
\begin{equation}
    (\delta(\varepsilon_i - \varepsilon_f + \omega))^2(\delta(p_{zi} - p_{zf} + k_z))^2 = \frac{T}{2\pi}\delta(\varepsilon_i - \varepsilon_f + \omega)\frac{L}{2\pi}\delta(p_{zi} - p_{zf} + k_z).
\end{equation}
To obtain the total absorption probability, $W$, one should take sums over the quantum numbers $s_f$ and $j_f$ of the final electron state and integrate over the final electron momentum projection, $p_{zf}$. The corresponding total absorption probabilities for various spin transitions of an electron are
\begin{multline} \label{eq:Wuu}
    W_{\uparrow\uparrow} = \frac{T}{3\omega LR} 4 e^2 N_i^{\uparrow 2} N_f^{\uparrow 2} L^2 \pi^4 \kappa \rho_H^4 \left(\sqrt{(\varepsilon_i + \omega)^2 + \frac{4}{\rho_H^2}} + \varepsilon_i + \omega\right) \Bigg[2\sqrt{2}\Bigg((\varepsilon_i + m_e)d^{(1)}_{-1\Lambda}(\theta_k)\mathcal{F}_{s_f,s_i}^{j_i + m + 1/2,j_i - 1/2}(\kappa\rho_H) - \\ - (\varepsilon_i + \omega + m_e)d_{1\Lambda}^{(1)}(\theta_k)\mathcal{F}_{s_f,s_i}^{j_i + m - 1/2, j_i + 1/2}(\kappa\rho_H)\Bigg) + \\
     + \rho_H d^{(1)}_{0\Lambda}(\theta_k) \left(m_e(2p_{zi} + k_z) + (p_{zi} + k_z)\varepsilon_i + p_{zi}(\varepsilon_i + \omega)\right)\mathcal{F}_{s_f,s_i}^{j_i + m - 1/2,j_i - 1/2}(\kappa\rho_H)\Bigg]^2, 
\end{multline}
\begin{multline} \label{eq:Wdu}
    W_{\downarrow\uparrow} = \frac{T}{3\omega LR} 8 e^2 N_i^{\uparrow 2} N_f^{\downarrow 2} L^2 \pi^4 \kappa \rho_H^4 \left(\sqrt{(\varepsilon_i + \omega)^2 + \frac{4}{\rho_H^2}} + \varepsilon_i + \omega\right)\times \\ 
    \times \Bigg[\rho_H \left(-m_e k_z - (p_{zi} + k_z)\varepsilon_i + p_{zi}(\varepsilon_i + \omega)\right)d_{-1\Lambda}^{(1)}(\theta_k)\mathcal{F}_{s_f,s_i}^{j_i + m + 1/2,j_i - 1/2}(\kappa\rho_H) + \\
    + \sqrt{2} d_{0\Lambda}^{(1)}(\theta_k)\left((\varepsilon_i + m_e)(s_f + j_i + m + 1/2)\mathcal{F}_{s_f,s_i}^{j_i + m - 1/2,j_i - 1/2}(\kappa\rho_H) - (\varepsilon + \omega + m_e)\mathcal{F}_{s_f,s_i}^{j_i + m + 1/2, j_i + 1/2}(\kappa\rho_H)\right)\Bigg]^2,
\end{multline}
\begin{multline} \label{eq:Wud}
    W_{\uparrow\downarrow} = \frac{T}{3\omega LR} 8 e^2 N_i^{\downarrow 2} N_f^{\uparrow 2} L^2 \pi^4 \kappa \rho_H^4 \left(\sqrt{(\varepsilon_i + \omega)^2 + \frac{4}{\rho_H^2}} + \varepsilon_i + \omega\right) \times \\ 
    \times \Bigg[\rho_H \left(-m_e k_z - (p_{zi} + k_z)\varepsilon_i + p_{zi}(\varepsilon_i + \omega)\right)d_{1\Lambda}^{(1)}(\theta_k)\mathcal{F}_{s_f,s_i}^{j_i + m - 1/2,j_i + 1/2}(\kappa\rho_H) + \\
    + \sqrt{2} d_{0\Lambda}^{(1)}(\theta_k)\left((s_i + j_i + 1/2)(m_e + \varepsilon + \omega) \mathcal{F}_{s_f,s_i}^{j_i + m - 1/2, j_i - 1/2}(\kappa\rho_H) - (m_e + \varepsilon) \mathcal{F}_{s_f,s_i}^{j_i + m + 1/2, j_i + 1/2}(\kappa\rho_H)\right)\Bigg]^2,
\end{multline}
\begin{multline} \label{eq:Wdd}
    W_{\downarrow\downarrow} = \frac{T}{3\omega LR} 4 e^2 N_i^{\downarrow 2} N_f^{\downarrow 2} L^2 \pi^4 \kappa \rho_H^4 \left(\sqrt{(\varepsilon_i + \omega)^2 + \frac{4}{\rho_H^2}} + \varepsilon_i + \omega\right) \times \\
    \times \Bigg[2\sqrt{2}(s_i + j_i + 1/2)(m_e + \varepsilon_i + \omega)d_{-1\Lambda}^{(1)}(\theta_k)\mathcal{F}_{s_f,s_i}^{j_i + m + 1/2,j_i - 1/2}(\kappa\rho_H) - \\ - 2\sqrt{2}(s_f + j_i + m + 1/2)(m_e + \varepsilon)d_{1\Lambda}^{(1)}(\theta_k)\mathcal{F}_{s_f,s_i}^{j_i + m - 1/2,j + 1/2}(\kappa\rho_H) + \\ + \rho_H d_{0\Lambda}^{(1)}(\theta_k)\left(m_e(2p_{zi} + k_z) + (p_{zi} + k_z)\varepsilon_i + p_{zi}(\varepsilon_i + \omega)\right)\mathcal{F}_{s_f,s_i}^{j_i + m + 1/2,j_i + 1/2}(\kappa\rho_H)\Bigg]^2.
\end{multline}

\section{Generalized cross section}
\label{app:cross_sect}
To analyze the absorption process in detail, we use the concept of generalized cross sections, which applies when the initial particles are wave packets instead of plane waves. This idea was developed in works such as \cite{Kotkin1992, Karlovets2020Apr, Karlovets2017Mar}. We need to introduce this concept because the absorption probabilities include normalization factors, $L$ and $R$, which don’t cancel out in the final results. This is similar to plane wave scattering, where the key quantity, independent of the normalization volume, is the ratio of the scattering probability to the incident particle flux. Likewise, the generalized cross section is defined as the ratio of the process probability \eqref{eq:diff_prob} to the luminosity.
\begin{equation} \label{eq:diff_sig}
    d\sigma = \frac{dW}{\mathcal{L}}.
\end{equation}
The luminosity is given by the following expression
\begin{equation} \label{eq:lum}
    \mathcal{L} = \int d^4x \frac{d^3p}{(2\pi)^3}\frac{d^3k}{(2\pi)^3} v(\bm{p},\bm{k}) w^{(e)}(\bm{r},\bm{p}) w^{(\gamma)}(\bm{r},\bm{k}),
\end{equation}
where $w^{(e)}(\bm{r},\bm{p})$ and $w^{(\gamma)}(\bm{r},\bm{k})$ are the electron and photon Wigner functions respectively, and
\begin{equation}
    v(\bm{p},\bm{k}) = \frac{p_{\mu}k^{\mu}}{\varepsilon(\bm{p})\omega(\bm{k})} = 1 - \frac{\bm{p}\cdot\bm{k}}{\varepsilon(\bm{p})\omega(\bm{k})}
\end{equation} is the relative velocity of the plane wave components of the initial electron and incident photon. When the initial particles are plane waves, the luminosity is calculated by multiplying the relative velocity with the probability densities of the particles. It’s important to note that in equation \eqref{eq:lum}, the Wigner functions are not assumed to be positive everywhere. A detailed explanation of the Wigner functions for electrons and photons, along with the explicit formula for luminosity, is provided in Appendix \ref{luminosity}.

\section{Appendix: Luminosity}
\label{luminosity}
The luminosity of the initial electron and incident photon according to \cite{Kotkin1992}, \cite{Karlovets2020Apr} and \cite{Karlovets2017Mar} is defined as
\begin{equation}
    \mathcal{L} = \int d^4x\frac{d^3p}{(2\pi)^3}\frac{d^3k}{(2\pi)^3}v(\bm{p},\bm{k})w^{(e)}_{s,j\mp 1/2}(\bm{r},\bm{p},t)w^{(\gamma)}(\bm{r},\bm{k},t),
\end{equation}
where
\begin{equation}
    v(\bm{p},\bm{k}) = \frac{p_{\mu}k^{\mu}}{\varepsilon(\bm{p})\omega(\bm{k})} = \frac{\varepsilon(\bm{p})\omega(\bm{k}) - \bm{p}\cdot{\bm{k}}}{\varepsilon(\bm{p})\omega(\bm{k})} = 1 - \frac{\bm{p}\cdot\bm{k}}{\varepsilon(\bm{p})\omega(\bm{k})}
\end{equation}
is the relative velocity between the plane wave components of the initial electron and incident photon and
\begin{equation}
    \begin{aligned}
        & w^{(e)}_{s,j\mp 1/2}(\bm{r},\bm{p},t) = \int \Psi_{s,j\mp 1/2}^{\dagger}(\bm{r}-\bm{y}/2,t)\Psi_{s,j\mp 1/2}(\bm{r} + \bm{y}/2,t)e^{i\bm{p}\cdot\bm{r}}d^3\bm{y},\\
        & w^{(\gamma)}(\bm{r},\bm{p},t) = \frac{1}{\omega}\int \bm{E}_{\kappa m k_z \Lambda}^{*}(\bm{r}-\bm{y}/2,t)\bm{E}_{\kappa m k_z \Lambda}(\bm{r} + \bm{y}/2,t)e^{i\bm{k}\cdot\bm{r}}d^3\bm{y}
    \end{aligned}
\end{equation}
are the Wigner functions of the initial electron and incident photon respectively. Here $\bm{E}_{\kappa m k_z \Lambda}$ is the electric field of the incident photon. The Wigner functions are defined in a way that when integrated over momentum space ($d^3k/(2\pi)^3$ for photons and $d^3p/(2\pi)^3$ for electrons), they give the spatial distribution of the particle number:
\begin{equation}
\label{eq:wig_to_probdist}
    \begin{aligned}
        & \int\frac{d^3p}{(2\pi)^3}w_{s, j \mp 1/2}^{(e)}(\bm{r},\bm{p},t) = |\Psi_{s,j \mp 1/2}(\bm{r},t)|^2;\\
        & \int\frac{d^3k}{(2\pi)^3}w^{(\gamma)}(\bm{r},\bm{k},t) = \frac{|\bm{E}(\bm{r},t)|^2}{\omega} = \omega |\bm{A}(\bm{r},t)|^2.
    \end{aligned}
\end{equation}
Let us rewrite the luminosity corresponding to the ``spin-up'' and ``spin-down'' electron states as follows
\begin{equation} \label{eq:lum_u}
    \mathcal{L}_\uparrow = 2\pi N_{s,j - 1/2}\left(\rho^2_H ((\varepsilon + m_e)^2 + p_{0z}^2)\ell_+ + 4\ell_-\right),
\end{equation}
\begin{equation} \label{eq:lum_d}
    \mathcal{L}_\downarrow = 2\pi N_{s,j + 1/2}\left(\rho^2_H ((\varepsilon + m_e)^2 + p_{0z}^2)\ell_- + 4(s + j + 1/2)^2\ell_+\right),
\end{equation}
where $N_{s,j\mp 1/2}$ is a normalization constant given by \eqref{eq:e_norm_const}, and the quantities $\ell_\pm$ are
\begin{equation}
    \ell_{\pm} = \int d^4x\frac{d^3p}{(2\pi)^3}\frac{d^3k}{(2\pi)^3}v(\bm{p},\bm{k})n_{s,j \mp 1/2}^{(e)}(\bm{r},\bm{p},t)w^{(\gamma)}(\bm{r},\bm{k},t).
\end{equation}
Here $n_{s,j \mp 1/2}^{(e)}(\bm{r},\bm{p},t)$ is the reduced electron Wigner function:
\begin{equation}
    n_{s,j \mp 1/2}^{(e)}(\bm{r},\bm{p},t) = \int \psi_{s,j\mp 1/2}^*(\bm{r}-\bm{y}/2,t)\psi_{s,j\mp 1/2}(\bm{r} + \bm{y}/2,t)e^{i\bm{p}\cdot\bm{r}}d^3y
\end{equation}
where $\psi_{s,j\mp 1/2}(\mathbf{r})$ is defined by \eqref{sol}. Notice that we can write
\begin{equation} \label{eq:ell12_not_exp}
\begin{aligned}
    &\ell_\pm = \int d^4x \left(\int\frac{d^3p}{(2\pi)^3}n_{s, j \mp 1/2}^{(e)}(\bm{r},\bm{p},t)\right)\left(\int\frac{d^3k}{(2\pi)^3}w^{(\gamma)}(\bm{r},\bm{k},t)\right) - \\
    &\int d^4x \int \frac{d^3p}{(2\pi)^3}\frac{d^3k}{(2\pi)^3}\frac{\bm{p}\cdot\bm{k}}{\varepsilon(\bm{p})\omega(\bm{k})}n_{s,j \mp 1/2}^{(e)}(\bm{r},\bm{p},t)w^{(\gamma)}(\bm{r},\bm{k},t).
\end{aligned}
\end{equation}
The first term here can be factored and rewritten in terms of the wave functions accroding to \eqref{eq:wig_to_probdist}, while the second term includes the dot product of the electron and photon momenta, which can be expanded as follows
\begin{equation}
    \bm{p}\cdot\bm{k} = pk\cos(\varphi_p - \varphi_k) + p_zk_z = pk\cos(\varphi_p)\cos(\varphi_k) + pk\sin(\varphi_p)\sin(\varphi_k) + p_zk_z.
\end{equation}
One can rewrite \eqref{eq:ell12_not_exp} in the following way
\begin{equation}
\label{lumin}
\begin{aligned}
    &\ell_\pm = \int d^4x |\psi_{s, j \mp 1/2}(\bm{r},t)|^2\frac{|\bm{E}(\bm{r},t)|^2}{\omega} - \\
    &\int d^4x \left(\int\frac{d^3p}{(2\pi)^3}\frac{p\cos(\varphi_p)}{\varepsilon(\bm{p})}n_{s, j \mp 1/2}^{(e)}(\bm{r},\bm{p},t)\right)\left(\int\frac{d^3k}{(2\pi)^3}\frac{k\cos(\varphi_k)}{\omega(\bm{k})}w^{(\gamma)}(\bm{r},\bm{k},t)\right) -\\
    &\int d^4x \left(\int\frac{d^3p}{(2\pi)^3}\frac{p\sin(\varphi_p)}{\varepsilon(\bm{p})}n_{s, j \mp 1/2}^{(e)}(\bm{r},\bm{p},t)\right)\left(\int\frac{d^3k}{(2\pi)^3}\frac{k\sin(\varphi_k)}{\omega(\bm{k})}w^{(\gamma)}(\bm{r},\bm{k},t)\right)-\\
    &\int d^4x \left(\int\frac{d^3p}{(2\pi)^3}\frac{p_z}{\varepsilon(\bm{p})}n_{s, j \mp 1/2}^{(e)}(\bm{r},\bm{p},t)\right)\left(\int\frac{d^3k}{(2\pi)^3}\frac{k_z}{\omega(\bm{k})}w^{(\gamma)}(\bm{r},\bm{k},t)\right).
\end{aligned}
\end{equation}
Here, the second and third terms vanish due to the integration of cosine and sine functions over the whole azimuthal angle range, $[0,2\pi]$, which in turn results in the simplification
\begin{equation} \label{eq:ell_12}
    \begin{aligned}
    &\ell_\pm = \int d^4x |\psi_{s, j \mp 1/2}(\bm{r},t)|^2\frac{|\bm{E}(\bm{r},t)|^2}{\omega} - \\
    &p_{0z}k_{0z}\int d^4x \left(\int\frac{d^3p}{(2\pi)^3}\frac{1}{\varepsilon(\bm{p})}n_{s, j \mp 1/2}^{(e)}(\bm{r},\bm{p},t)\right)\left(\int\frac{d^3k}{(2\pi)^3}\frac{1}{\omega(\bm{k})}n^{(\gamma)}(\bm{r},\bm{k},t)\right).
    \end{aligned}
\end{equation}
Note that, since both electron and photon states possess a definite value of their longitudinal momenta, their Wigner functions include these quantities only in delta functions $\delta(p_z - p_{0z})$ and $\delta(k_z - k_{0z})$ for given $p_{0z}$ and $k_{0z}$. Therefore, in the last expression the values $p_z$ and $k_z$ can be replaced by $p_{0z}$ and $k_{0z}$ respectively. Since the luminosity is a Lorenz invariant quantity, one can write it in the frame, where $p_{0z} = 0$. In this frame the second term in \eqref{eq:ell_12} vanishes and $\ell_\pm$ simplifies to
\begin{equation} \label{eq:ell_12_rf}
    \ell_\pm = \omega\int d^4x |\psi_{s, j \mp 1/2}(\bm{r},t)|^2 |\bm{A}(\bm{r},t)|^2 = 2\pi TL\omega \int d\rho\,\rho |\psi_{s, j \mp 1/2}(\bm{r},t)|^2 |\bm{A}(\bm{r},t)|^2.
\end{equation} 
This representation of $\ell_\pm$ simplifies the process by only requiring the integration of the spatial distribution functions, without the need for explicitly calculating the Wigner functions, which are quite complex to compute.
\clearpage

\bibliographystyle{apsrev}
\bibliography{references}

\end{document}